\documentclass[10pt,nofootinbib,aps,pra]{revtex4-1}

\usepackage{latexsym}
\usepackage{amsmath}
\usepackage{amsfonts}

\usepackage{hhline}
\usepackage{graphicx}
\usepackage{amssymb}

\usepackage{multirow}
\usepackage{rotating}

\usepackage{comment}

\usepackage[
        plainpages=false, 
        pdfpagelabels, 
        bookmarks=true, 
        bookmarksnumbered = true, 
        bookmarksopen=false, 
        bookmarksopenlevel=3, 
        colorlinks=true,
        citecolor=blue,
        urlcolor=black,
        linkcolor=black,
        ]{hyperref}

\newcommand{\unit}[1]{\ensuremath{\, \mathrm{#1}}}

\begin{document}
\title{Understanding the dynamics of rings in the melt in terms of annealed tree model}
\author{Jan Smrek}
\email{js5013@nyu.edu}
\author{Alexander Y. Grosberg}
\affiliation{Center for Soft Matter Research and Department of Physics, New York
University, New
York, NY 10003, USA}
\begin{abstract}
Dynamical properties of a long polymer ring in a melt of unknotted and unconcatenated rings are calculated. We re-examine and generalize the well known model of
a ring confined to a lattice of topological obstacles in the light of the recently developed Flory theory of unentangled rings which maps every ring on an
annealed branched polymer and establishes that the backbone associated with each ring follows self-avoiding rather than Gaussian random walk statistics. We find
the scaling of ring relaxation time and diffusion coefficient with ring length, as well as time dependence of stress relaxation modulus, zero shear viscosity
and mean square averaged displacements of both individual monomers and ring's mass center. Our results agree within error bars with all available experimental
and simulations data of the ring melt, although the quality of the data so far is insufficient to make a definitive judgment for or against the annealed tree
theory. In the end we review briefly the relation between our 
findings and 
experimental data on chromatin dynamics.
\end{abstract}
\maketitle

\nocite{*}
\section{Introduction}
Recent progress in experimental studies of genome 3D organization in eukaryotic nuclei (see e.g.
\cite{HiC_experiments_review_2012,Dekker_Chapter_7,HiC_Science_2009}) along with earlier theoretical suggestions
\cite{crumpled1,crumpled2,Rosa_Everaers_PLOS_2008} shed light on the delicate interplay of high density and topological constraints in genome 3D structure and
dynamics. This interplay certainly belongs to the most challenging and least understood aspects of polymer physics \cite{Crumpled_globule_Grosberg_review,
Mirny_one_ring_2014, Melt_of_rings_Statics_2011, Rosa_Orlandini_Macromolecules_2011}. While there are still many unanswered questions related to the statics and
the structure of dense topologically constrained polymer matter, their dynamics is even more difficult. In this paper we consider the simplest problem of this
type. We claim our problem to be the simplest because of two decisive simplifications: The first simplification is we examine prototypical system exhibiting
competition between high density and topological constraints -- the melt of unknotted and unconcatenated rings. The relation of this model to genome folding was
discussed at length (see e.g. \cite{Rosa_Everaers_PLOS_2008,Crumpled_globule_Grosberg_review,Crumpled_globule_Mirny_review,Crumpled_globule_Mirny_review_2}, so
here we do not touch on that. The relevance of our dynamics results and their potential to clarify experiments on chromatin dynamics is discussed at the very
end of our paper. In the main text we work on the model system -- on passive dynamics of rings.

\begin{figure}[htb]
 \includegraphics[width=0.4\columnwidth]{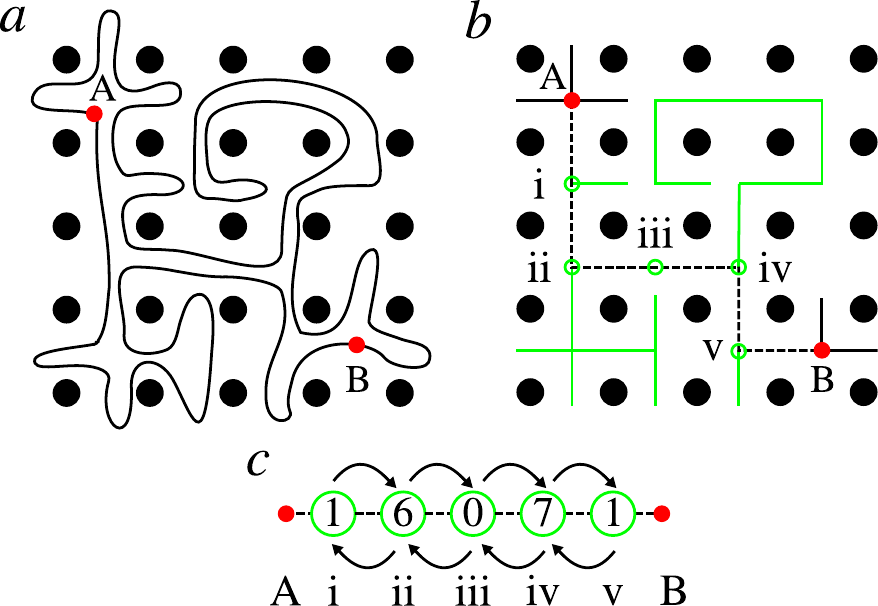}
 \caption{$(a)$ Ring conformation (here shown in 2D for simplicity) in the lattice of topological obstacles (black filled circles) $(b)$ The corresponding tree
conformation (on dual lattice). Each segment represents doubly folded ring part. Dashed line A-i-ii-iii-iv-v-B is a simplistic representation of the backbone
$L$ (see Appendix \ref{app_L_definition} for details). Each lattice point on the backbone represents a reservoir (green empty circle) of blobs of stored length
that form the branches (green solid lines). Thermal motion of the ring is modeled as the diffusion of the blobs from one branch to another. As the ring is
relaxing this happens on progressively larger length scale. Here we show for simplicity the largest length scale only. $(c)$ Reservoir representation of the
blob diffusion. For the specific conformation shown in figure $b$ the number of stored blobs for each reservoir is shown.}
 \label{fig_conformation}
\end{figure}

Our second crucial simplification comes directly in the context of ring melt: we will construct the dynamical description based on the recently proposed static
Flory-like theory \cite{Grosberg_ring_melt_static}. The origin of that theory goes back to works
\cite{Rubinstein_PRL_1986,Rubinstein_ring_in_gel_dynamics,Khokhlov_Nechaev_1985,Nechaev_lattice_of_obstacles}. As a reminder, the work
\cite{Khokhlov_Nechaev_1985} considered a single loop, having no excluded volume and placed in a lattice of topological obstacles, such that the loop is not
threaded by any of the obstacles (see Fig.~\ref{fig_conformation} $a$). In this case, the ring adopts a branched conformation of a tree (also called a lattice
animal in this context), with each branch representing a doubly folded section of the ring (Fig.~\ref{fig_conformation} $b$); it was shown
\cite{Khokhlov_Nechaev_1985}, that this tree is randomly branched. In
\cite{Nechaev_lattice_of_obstacles,Rubinstein_PRL_1986,Rubinstein_ring_in_gel_dynamics,Turner_ring_selfthreading_gel} authors examined the dynamics of this
system.

Since randomly branched tree has gyration radius $R$ scaling as $N^{1/4}$, $N$ being polymerization index (see e.g. \cite{Rubinstein_book}), these results do
not directly apply to the real rings with excluded volume in 3D (and even more so in 2D), unless $N$ is rather small. Nevertheless, authors of the work
\cite{Khokhlov_Nechaev_1985} made a very far-sighted comment at the end of their article suggesting that ring compaction in the lattice of obstacles is an
indication of globular conformations (with gyration radius $\sim N^{1/3}$) of the real rings in an unconcatenated melt. Subsequent developments, particularly
recent large scale simulations \cite{Melt_of_rings_Statics_2011, Rosa_Everaers_PRL_2014} seem to have accepted this idea of compactness ($R\sim N^{1/3}$) for
each ring in an unconcatenated melt. Flory-type theory \cite{Grosberg_ring_melt_static} was set to rationalize this finding.

To do so, the work \cite{Grosberg_ring_melt_static} assumed in line with
\cite{Khokhlov_Nechaev_1985,Nechaev_lattice_of_obstacles,Rubinstein_PRL_1986,Rubinstein_ring_in_gel_dynamics}, that crumpled conformation of a ring in
an unconcatenated melt can be mapped on a tree, with two important modifications
against \cite{Khokhlov_Nechaev_1985,Nechaev_lattice_of_obstacles,Rubinstein_PRL_1986,Rubinstein_ring_in_gel_dynamics}: first, tree branches do not have to
be doubly folded down to the monomer scale, they may include (and likely to include) loops; second, the tree is not assumed random, rather its branches
are viewed as \emph{annealed}, subject to thermal equilibration. The latter is described by the special order parameter denoted as $L$ and
called \emph{backbone}.

An important piece of evidence in support for the existence of an underlying tree structure for every ring in the melt was provided in recent simulation work
\cite{Rosa_Everaers_PRL_2014}. There, authors have demonstrated quantitatively accurate mapping of rings on properly chosen trees. We should also mention that
tree-like structures were implicated in chromatin context in the work \cite{Iyer_Arya_2012}.

To make present paper self-contained, we include in Appendix \ref{app_L_definition} further discussion of $L$, its meaning and definition. For the same purpose,
we discuss in Appendix \ref{app_mapping} the (simpler case of) analogous theory in 2D, where Flory theory is, as usually, not exact but very accurate.

In this paper, our plan is to use methods of the work \cite{Rubinstein_ring_in_gel_dynamics} to analyze the dynamics of rings viewed as annealed, but not ideal
branched objects. In \cite{Rubinstein_ring_in_gel_dynamics}, the ideal branched object statistics was considered. The backbone of the ring was assumed to behave
as a random walk in space implying the gyration radius $R$ scaling with the backbone $L$ as $R\sim L^{1/2}$. The length $L$ itself was assumed to scale with the
number of monomers of the ring $N$ also with exponent $1/2$ i.e. $L\sim N^{1/2}$, which gives the exponent $\nu$ governing the spatial size of the ring
(gyration radius) $R\sim N^{\nu}$ to be $\nu=1/4$ \cite{Zimm_Stockmayer}. By contrast, Flory-type theory \cite{Grosberg_ring_melt_static} establishes that rings
in the melt represent annealed but not ideal trees, with $R\sim N^{1/3}$ i.e. $\nu=1/3$ while the backbone scales as $L\sim N^{\rho}$ with
$\rho=1/3\nu_{\mathrm{F}}\simeq 0.567$ where $\nu_{\mathrm{F}}=0.588$ is the Flory exponent of regular self-avoiding walk in 3D. The exponents $\nu=1/3$ and
$\rho\simeq0.567$ represent compact ring conformations with backbone following a self-avoiding random walk statistics as is shown below. The goal of this paper
is to reproduce the main dynamics arguments of \cite{Rubinstein_ring_in_gel_dynamics} for this new exponents $\rho$ and $\nu$ and to compare it with the most
detailed computer data available \cite{Kremer_rings_dynamics}.

\section{Calculation details}
\label{sec_calc}
Here, we derive the time dependence of the stress relaxation modulus, mean square displacements of monomers and center of mass and the behavior of the zero
shear viscosity as function of the ring mass. At first, we briefly summarize the conformational properties of the rings with the annealed tree statistics from
\cite{Grosberg_ring_melt_static} and review the ring relaxation dynamics of \cite{Rubinstein_ring_in_gel_dynamics}. Later in section \ref{sec_results} we show
how these theoretical predictions compare with experiments \cite{Kapnistos_Rubinstein_experimental,Vlassopoulos_viscosity_rings_2013} and extensive numerical
simulations \cite{Kremer_rings_dynamics} and in section \ref{sec_discussion} we discuss relation to dynamics of eukaryotic DNA.

Following \cite{Grosberg_ring_melt_static}, ring is a annealed tree of ``entanglement'' blobs each consisting of about $N_{e}$ monomers, where the entanglement
length $N_{e}$ is the crossover between Gaussian and compact regime $\nu=1/3$ for the ring gyration radius $R$. 
\begin{displaymath}
 R\sim\left\{ \begin{array}{lll}         
	      N^{1/2} 
&\textrm{for}& N<N_{e} \\
	      N^{1/3}N_{e}^{1/6} &\textrm{for}& N>N_{e} 
             \end{array} \right.
\end{displaymath}
The monomer size and Kuhn segment are assumed for simplicity to be of the same order and are taken as a unit of length. The gyration radius of the Gaussian
``entanglement'' blob $N_{e}^{1/2}$ sets the lattice constant of the topological mesh formed by other rings in the melt. As mentioned above, the backbone of the
annealed tree $L$ scales with the ring length as $L\sim N^{\rho}$ with $\rho=1/3\nu_{\mathrm{F}}$ where $\nu_{\mathrm{F}}\simeq0.588$ is the Flory exponent of
regular self-avoiding walk in 3D. This says the backbone of the annealed tree is a self-avoiding random walk in space because $R\sim L^{\nu/\rho}\sim
L^{\nu_{\mathrm{F}}}$. To summarize, while the annealed tree conformations are characterized by exponents $\nu=1/3$ and $\rho\simeq 0.567$, the randomly
branched (ideal) trees are described by $\nu_{id}=1/4$ and $\rho_{id}=1/2$.

Theoretical work \cite{Rubinstein_ring_in_gel_dynamics} that we follow showed an important difference between relaxation of a ring and a linear polymer. The
ring relaxation time is dominated by the relaxation of its backbone as that is the most stable structure of the conformation. Diffusion of the blobs along the
backbone of the conformation is modeled as a diffusion of stored length between different branches (side loops) that serve as reservoirs of the blobs of stored
length (Fig. \ref{fig_conformation} $b$ and $c$). This approach tries to capture the change of the shape of the ring during a diffusion of a blob along the ring
caused by a diffusion of other blobs. 

We consider scaling estimates for all quantities examined in computational work \cite{Kremer_rings_dynamics}. Let's start with \underline{relaxation time} of
the entire ring. Displacement of a single blob to a neighboring reservoir at a distance one lattice constant, takes about a Rouse time $\tau_{0}\sim N_{e}^2
\zeta / kT$, where $\zeta$ is the fluid friction of a monomer. Each blob represents a fraction $N_{e}/N$ of the total mass of the ring, which means a single
blob displacement to a neighboring reservoir $N_{e}^{1/2}$ away causes the center of mass to move a distance $N_{e}^{1/2}/(N/N_{e})$. There are about
$(N/N_{e})^{\rho}$ reservoirs that exchange the blobs between themselves in time $\tau_{0}$. The different blob displacements are assumed mutually independent
(and non-interacting) hence the mean square displacement of the center of mass of a ring is $\Delta s_{\mathrm{cm}}^{2} \sim (N/N_{e})^{\rho}
N_{e}/(N/N_{e})^{2} = N_{e} (N/N_{e})^{\rho-2}$. These displacements with diffusion coefficient $D_{s}^{\mathrm{cm}}\sim\Delta s_{\mathrm{cm}}^{2}/\tau_{0}$
take place however along the backbone of the ring. The relaxation of the whole ring is achieved when the center of mass is displaced a distance of the order of
the whole backbone, that is in time $\tau\sim L^{2}/D_{s}^{\mathrm{cm}}$ which is found to scale as
\begin{equation}
\label{relaxation_time}
 \tau \sim \tau_{0} (N/N_{e})^{\rho + 2}
\end{equation}

From here the scaling of the \underline{diffusion coefficient} of the ring in real space is also derived in \cite{Rubinstein_ring_in_gel_dynamics}. In the
relaxation time $\tau$ the ring is displaced a distance of its own size  and we find $D\sim R^{2\nu}/\tau$ hence
\begin{equation}
\label{diff_coef}
 D \sim D_{e} (N/N_{e})^{2\nu - \rho - 2},
\end{equation}
where $D_{e}=N_{e}/\tau_{0}$ is the diffusion coefficient of the entanglement blob. 

To find the \underline{stress relaxation modulus} we use the standard single polymer dynamics approach \cite{Rubinstein_book}. Until time $\tau_{0}$ the
Gaussian blobs of $N_{e}$ monomers are relaxing by Rouse modes and the stress relaxation modulus decays inversely proportional to the square root of time. At
time $\tau_{0}$ the stress modulus is $kT$ per volume of entanglement blob i.e. $G(\tau_{0})\sim kT/N_{e}$. After $\tau_{0}$ the relaxation time $\tau_{p}$ of
the $p$-th mode can be calculated as a relaxation time of a ring of $N/N_{e} p$ segments due to self-similarity of the ring conformation, hence $\tau_{p} \sim
\tau_{0} (N/N_{e} p)^{\rho + 2}$. We can express the number of modes that relax at time $t$ as
\begin{equation}
 \label{mode_time_dependence}
p(t) \simeq (N/N_{e}) (\tau_{0}/t)^{1/(\rho+2)}. 
\end{equation}
At time $t$, segments of length $N/N_{e} p(t)$ relax and the stress relaxation modulus is $k T$ per volume occupied by such segment. 
\begin{equation}
\label{stress_relaxation_modulus_1}
G(t) N_{e} \sim k T\frac{N_{e}}{N} p(t) \sim k T \left(\frac{\tau_{0}}{t}\right)^{1/(\rho+2)}.
\end{equation}
 This holds until the whole ring relaxes ($t<\tau$) after which an exponential decay starts.
\begin{eqnarray}
\label{stress_relaxation_modulus_2}
G(t) N_{e}/ k T = \left\{ \begin{array}{ll}
 (t /\tau_{0})^{-1/2} & \textrm{if $t<\tau_{0}$}\\
(t/\tau_{0})^{-1/(\rho+2)} & \textrm{if $\tau_{0}<t<\tau $}\\
(t/\tau_{0})^{-1/(\rho+2)} e^{-t/\tau} & \textrm{if $t>\tau $}
\end{array} \right.
\end{eqnarray}
At the intermediate times there is no plateau modulus as for the linear polymer melts, but the stress modulus decays as power-law. This is because the
relaxation of a ring continues in a self-similar manner above the entanglement length, while linear chain relaxes only after it reptates out of the entanglement
tube. The separation of timescales in the linear case, where nothing can be said to relax on the time scale intermediate to $\tau_{0}$ and reptation time, is
responsible for the plateau modulus.

\underline{Zero-shear viscosity} can be calculated from the stress relaxation modulus as $\eta_{0} = \int_{0}^{\infty} G(t) dt$ which is up to numerical factor
equal to
\begin{equation}
\label{viscosity}
\eta_{0} \simeq G(\tau)\tau  \sim \frac{kT}{N_{e}} (N/N_{e})^{\rho+1} \tau_{0}.
\end{equation}

Three different types of \underline{mean-square displacement} (MSD) are computed in \cite{Kremer_rings_dynamics}:
\begin{enumerate}
 \item the MSD averaged over all monomers of a ring 

$g_{1}(t) = \langle|\mathbf{r}_{i}(t) - \mathbf{r}_{i}(0)|^{2}\rangle$
 \item MSD of monomers with respect to the center of mass 

$g_{2}(t)=\langle|\mathbf{r}_{i}(t) - \mathbf{r}_{\mathrm{cm}}(t) - \mathbf{r}_{i}(0) + \mathbf{r}_{\mathrm{cm}}(0)|^{2}\rangle$
 \item MSD of the center of mass 

$g_{3}(t) = \langle|\mathbf{r}_{\mathrm{cm}}(t) - \mathbf{r}_{\mathrm{cm}}(0)|^{2}\rangle$.
\end{enumerate}

Let's consider $g_{1}(t)$ first. The Rouse relaxation, characterized by $g_{1}(t)\sim t^{1/2} $ takes place up to the time $\tau_{0}$ where blobs of  size
$N_{e}^{1/2}$ are relaxed. Next, at time $t$ segments of length $N/N_{e}p(t)$ relax and move a distance of the order of their size: $g_{1}(\tau_{p}) \sim N_{e}
\left(N/N_{e}p(t)\right)^{2\nu}$. Substituting for $p(t)$ from (\ref{mode_time_dependence}), we get
\begin{equation}
 g_{1}(t) \sim N_{e} \left(t/\tau_{0}\right)^{2\nu/(2+\rho)}.
\end{equation}
This continues until time $\tau$ when the whole ring is relaxed and ordinary diffusion sets off. Hence for the MSD averaged over all monomers we have
\begin{displaymath}
 g_{1}(t)/N_{e} \sim \left\{ \begin{array}{ll}
			 (t/\tau_{0})^{1/2} & \textrm{for $t<\tau_{0}$} \\
			 (t/\tau_{0})^{2\nu/(2+\rho)} & \textrm{for $\tau_{0}<t<\tau$}\\
			 (N/N_{e})^{2\nu} (t/\tau) & \textrm{for $t>\tau$}
                       \end{array}\right.,
\end{displaymath}
The prefactor $1/N_{e}$ ensures a smooth crossover between the intermediate and late regime.

Now consider the MSD of the center of mass $g_{3}$. On timescale $\tau_{p}$ a segment of length $N/N_{e}p$ moves a distance of the order of $(N/N_{e}p)^{\nu}$. 
There are about $N_{e} p$ of these segments in the ring hence the center of mass moves $(N/N_{e}p)^{\nu}/N_{e} p$ for each of them. Since these displacements
are independent of each other (the relaxation on longer scales is to come at later times), the total MSD is just
$g_{3}(t) \sim N_{e} p (N/N_{e}p)^{2 \nu}/(N_{e}p)^{2} \sim N^{2\nu} (N_{e} p(t))^{-2\nu-1}$. Using again the mode time dependence (\ref{mode_time_dependence})
we get
\begin{equation}
 N g_{3}(t)\sim (t/\tau_{0})^{(2\nu + 1)/(\rho + 2)}.
\end{equation}
Using this approach also for $t<\tau_{0}$ we get $N g_{3}(t) \sim t$ which is standard Rouse behavior at early times. Again after time $\tau$ the normal diffusion starts. The MSD of the center of mass follows
\begin{displaymath}
 N g_{3}(t)/N_{e}^{2} \sim \left\{ \begin{array}{ll}
			t/\tau_{0} & \textrm{for $t<\tau_{0}$}\\
			(t/\tau_{0})^{(2\nu + 1)/(\rho + 2)} & \textrm{for $\tau_{0}<t<\tau$}\\
			t/\tau & \textrm{for $t>\tau$}
                       \end{array}\right.,
\end{displaymath}
where the factor $N_{e}^{-2}$ guarantees the smooth crossover.

MSD $g_{2}$ is the difference between $g_{1}$ and $g_{3}$ and it crosses over from Rouse $t^{1/2}$ at early times to a constant at later times $t>\tau$.

\section{Results}
\label{sec_results}
\begin{figure}[htb]
 \includegraphics[width=0.5\columnwidth]{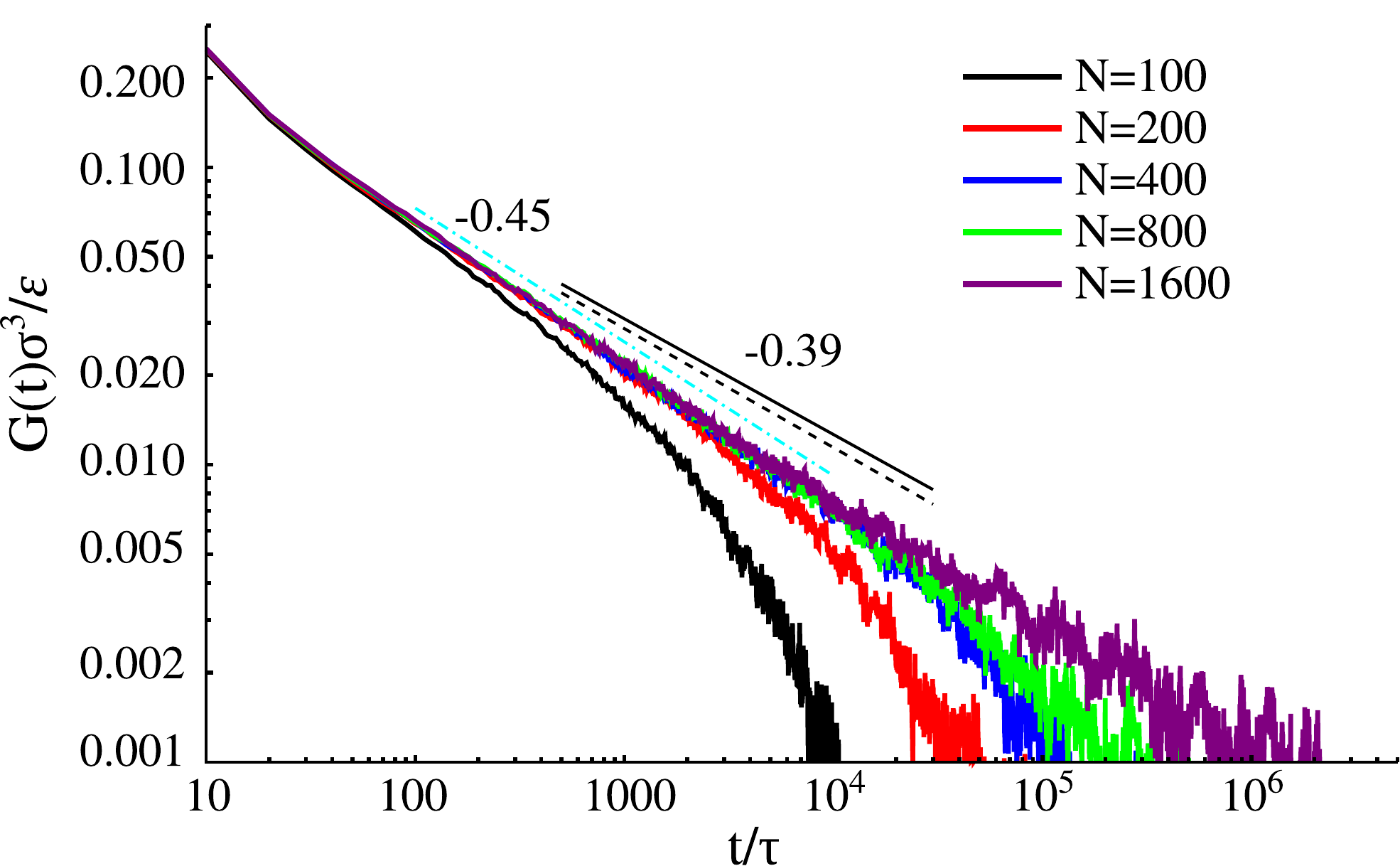}
 \caption{Relaxation modulus of rings of different $N$ as function of time. Straight solid black line has slope (eq.~\ref{stress_relaxation_modulus_2}) $-1/(\rho+2)\simeq-0.39$ (annealed tree), dashed line has slope $-0.4$ (ideal tree). The best fit of long rings data is cyan dot-dashed line with slope $-0.45$. $\sigma$ is the length scale and $\varepsilon$ is the energy scale of the interaction. See \cite{Kremer_rings_dynamics} for details.}
 \label{fig_relaxation_modulus}
\end{figure}

Let us now compare the results with available data. Experimentally it is difficult to purify melt of rings of linear contaminants that in sufficient amount have
strong effects on stress relaxation \cite{Kapnistos_Rubinstein_experimental}. Various studies agree that the critical concentration of the linear chains beyond
which the effects on viscosity of the melt become substantial lies well below the overlap concentration of the linear chains, but its exact value is still a
subject of active discussion \cite{Kapnistos_Rubinstein_experimental,Vasquez_simulation_contaminants_2011,Rings_Rheology_PRL2012}. Nevertheless, experimental
works \cite{Kapnistos_Rubinstein_experimental,Vlassopoulos_viscosity_rings_2013} on rings with less than $0.1\%$ of linear contaminants, confirmed the power-law
behavior of the stress relaxation modulus at intermediate times with power slightly below $-0.4$ in agreement with the numerical result of pure ring melt 
\cite{Kremer_rings_dynamics} of around $-0.45$. Let us stress that in this work we do not consider the problem of linear contaminants and the present theory
assumes pure melt of rings. Our calculation (\ref{stress_relaxation_modulus_1}) predicts the exponent $-1/(\rho+2)\simeq -0.390$ for the annealed tree, while
the ideal tree ($\rho_{id}=1/2$) gives $-0.4$ (See Fig.~\ref{fig_relaxation_modulus}).
While in the intermediate time regime, different experiments collapsed on the universal scaling curve, the single exponential cutoff at long times was not found
\cite{Vlassopoulos_viscosity_rings_2013}. This was attributed to the linear contaminants. 

\begin{figure}[htb]
 \includegraphics[scale=0.5]{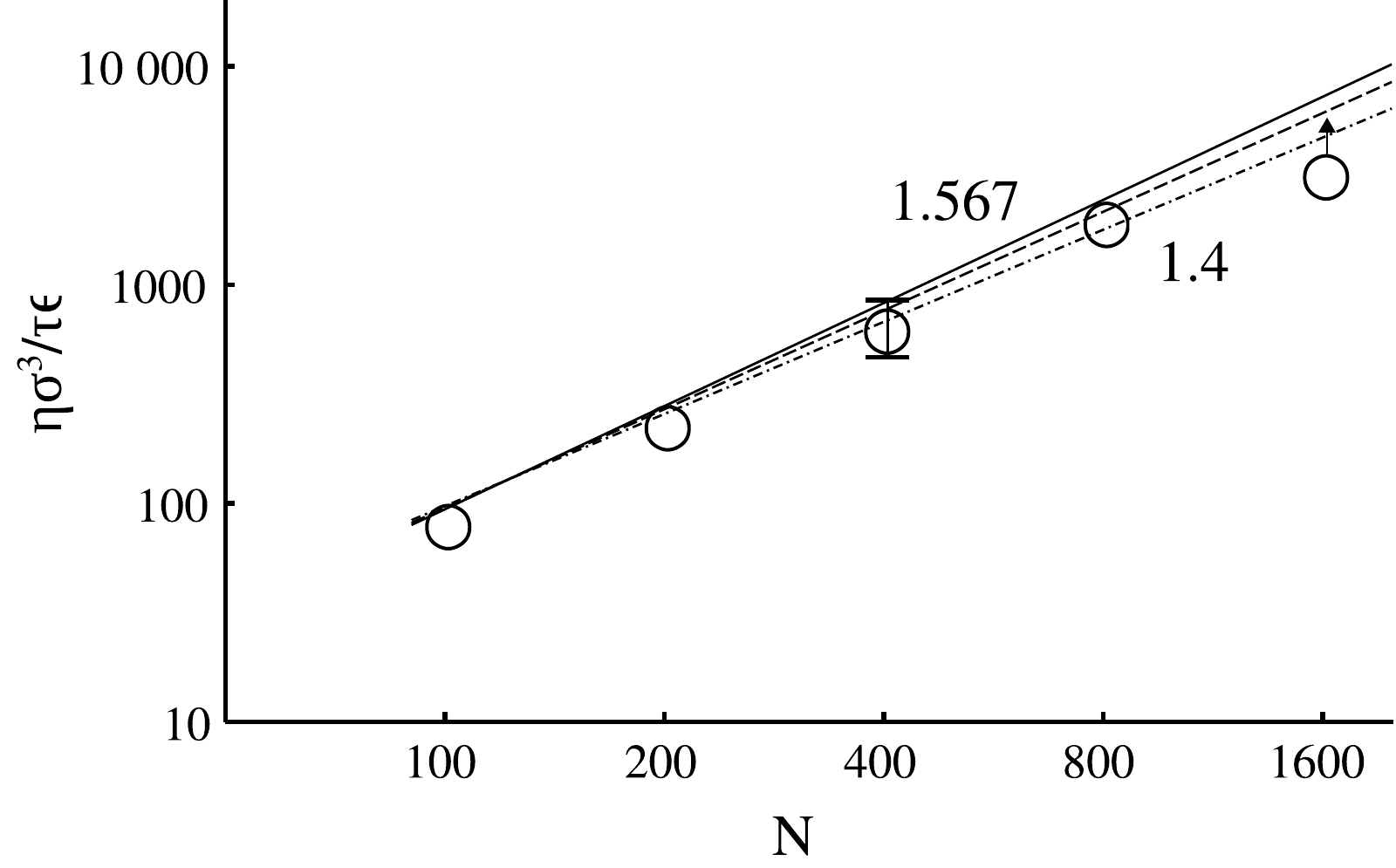}
 \caption{Zero-shear viscosity as function of $N$. Solid line has slope $1+\rho\simeq1.567$ (annealed trees theory), dashed line has slope $1.5$ (ideal trees),
dot-dashed line has slope $1.4$. Error bars for $N=400$ are from four independent simulations, other points are from single simulation. Error bars for $N=800$
are estimated to be smaller than twice the symbol size. Arrow at $N=1600$ points to extrapolated value of viscosity (from $t\to\infty$ limit of $g_{3}(t)$)
because a fully diffusive regime was not reached for rings of this length. See \cite{Kremer_rings_dynamics} for details.}
 \label{fig_viscosity}
\end{figure}

The experiments \cite{Vlassopoulos_viscosity_rings_2013} on ring melts report the ratio of zero-shear viscosity of linear melt to that of the rings
$\eta_{0,\textrm{linear}}/\eta_{0,\textrm{ring}}$ to scale with the number of entanglements $N/N_{e}$ with the exponent $1.2\pm 0.3$ for $N>N_{e}$. This is
smaller than the numerically measured exponent $2\pm0.2$ in \cite{Kremer_rings_dynamics}. There, the viscosity of linear melt was found to scale as
$\eta_{0,\textrm{linear}}\sim N^{3.4}$ in agreement with the reptation theory with tube length fluctuations corrections \cite{Rubinstein_book}. The ring melt
viscosity was found to scale as $\eta_{0,\textrm{ring}}\sim N^{1.4}$. This agrees with both predictions of the ring melt viscosity exponent (\ref{viscosity})
$1+\rho\simeq 1.567$ for self-avoiding backbone and $1.5$ for the ideal trees. See Fig.~\ref{fig_viscosity}.
If we take the exponent for the linear melt to be $3.4$ we predict $\eta_{0,\textrm{linear}}/\eta_{0,\textrm{ring}}$ to have the exponent $1.833$ and $1.9$ for
annealed and ideal trees respectively which is still higher than the reported experimental value. The lower exponents found in the experiments can be attributed
again to the linear contaminants. The linear contaminants would certainly increase the ring viscosity by threading the rings configuration, as noted in  
\cite{Vasquez_simulation_contaminants_2011,Vlassopoulos_viscosity_rings_2013,Rings_Rheology_PRL2012}, hence decreasing the exponent of
$\eta_{0,\textrm{linear}}/\eta_{0,\textrm{ring}}$. More experimental data is necessary to quantify the role of the contaminants and estimate the coefficient of
the pure melt of rings more accurately.

Numerical studies of a pure melt of rings \cite{Kremer_rings_dynamics,Rings_Rheology_PRL2012} provide and indispensable information for all the calculated
dynamical quantities. The numerical results for the diffusion coefficient give scaling $D\sim N^{-2.3}$ which is off by $0.4$ from our prediction
(\ref{diff_coef}) of the exponent $2\nu - \rho - 2 \simeq -1.9$. However, similar discrepancy between effective theory and experiment is found even for melts of
linear chains, where the reptation theory predicts $D\sim N^{-2}$ while experimental and numerical results are $D\sim N^{-2.4}$ \cite{Kremer_rings_dynamics}. In
the linear case the difference is attributed to the assumption that the topological obstacles are immobile. In reality, the constraints also move and relax due
to the motion of the other chains. To account for this effects, the reptation theory can be adjusted by contour length fluctuations or constraint release
\cite{Lodge_linear_melt_diffusion_coef} to give the exponents that agree with experiments. At the moment we don't know how to incorporate the relaxation of the
the topological constraints in the theory, but if the same corrections are assumed for the rings as found in the linear case, the simulation agrees well: in
both cases the theoretical prediction of the exponent is off by $0.4$ from the numerics. Using the exponents $\nu_{\mathrm{id}}=1/4$ and
$\rho_{\mathrm{id}}=1/2$ for the ideal trees, the exponent is $-2$ - the same as predicted by reptation theory for linear melts.

The most important numerical observation is that the diffusion coefficient for the rings and linear melts behaves (nearly) identically, but the viscosities
diverge dramatically. This suggests decoupling of the relaxation from the diffusion as the ring can relax without diffusing much. This is reproduced by the
present theory.

\begin{figure}[htb]
 \includegraphics[scale=0.4]{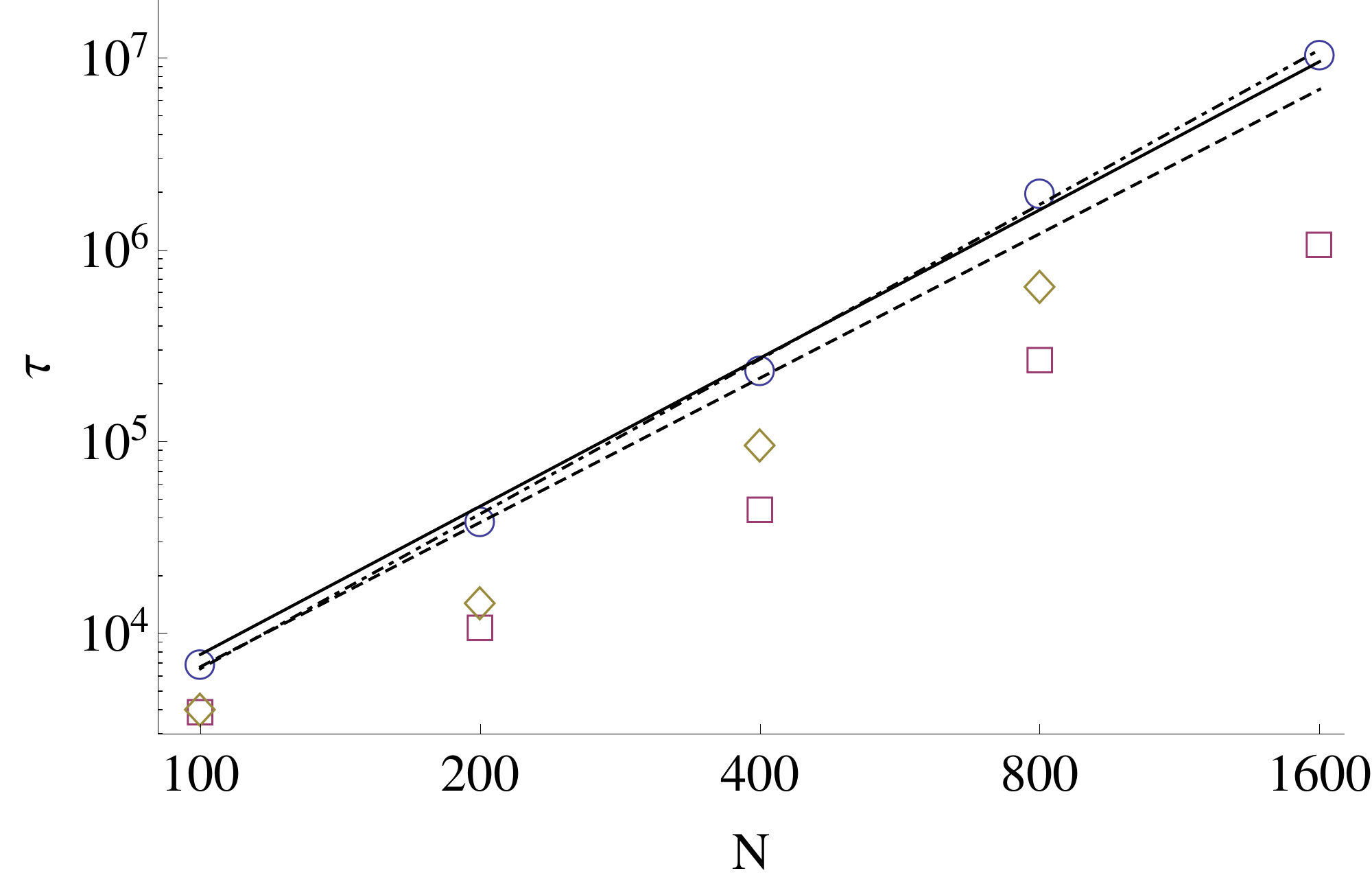}
 \caption{Different relaxation times as function of ring length $N$. Symbols are numerical results from \cite{Kremer_rings_dynamics}: diffusive relaxation time
(circles); conformational relaxation times: $\tau_{cc}$ (squares) represents decay of the correlation of two vectors connecting monomers $0$ with $N/2$ and
$N/4$ with $3N/4$ respectively - see details in \cite{Kremer_rings_dynamics}; $\tau_{R^{2}R^{2}}$ represents the decay of the $R^{2}$ autocorrelation function
(diamonds), solid line has slope $\rho+2 \simeq 2.567$ (annealed trees) by equation (\ref{relaxation_time}), dashed line has slope $5/2$ (ideal trees), while
the dot-dashed is the best fit of diffusive relaxation time with slope $2.68$.}
 \label{fig_relaxation_time}
\end{figure}

The numerical study presents three relaxation times to characterize the different aspects of relaxation. The diffusive relaxation time and the correlation time
of the gyration radius follow each other, while the internal rearrangement time is much smaller and also grows slightly more slowly with the mass of the ring
(see Figure \ref{fig_relaxation_time}). This can be understood in the context of the present theory as the internal rearrangement by blob diffusion around the
ring can take place without the ring diffusing too much. Our calculated exponents for the relaxation time, $2+\rho = 2.567$ for trees with self-avoiding
backbone and $2.5$ for ideal trees, are both slightly below the best fit exponent for the diffusive time $2.68$. The larger exponent for the trees with
self-avoiding backbone is a consequence of their less compact conformation due to self-avoidance hence the ring has to diffuse further to relax. 

\begin{figure}[htb]
 \includegraphics[scale=0.5]{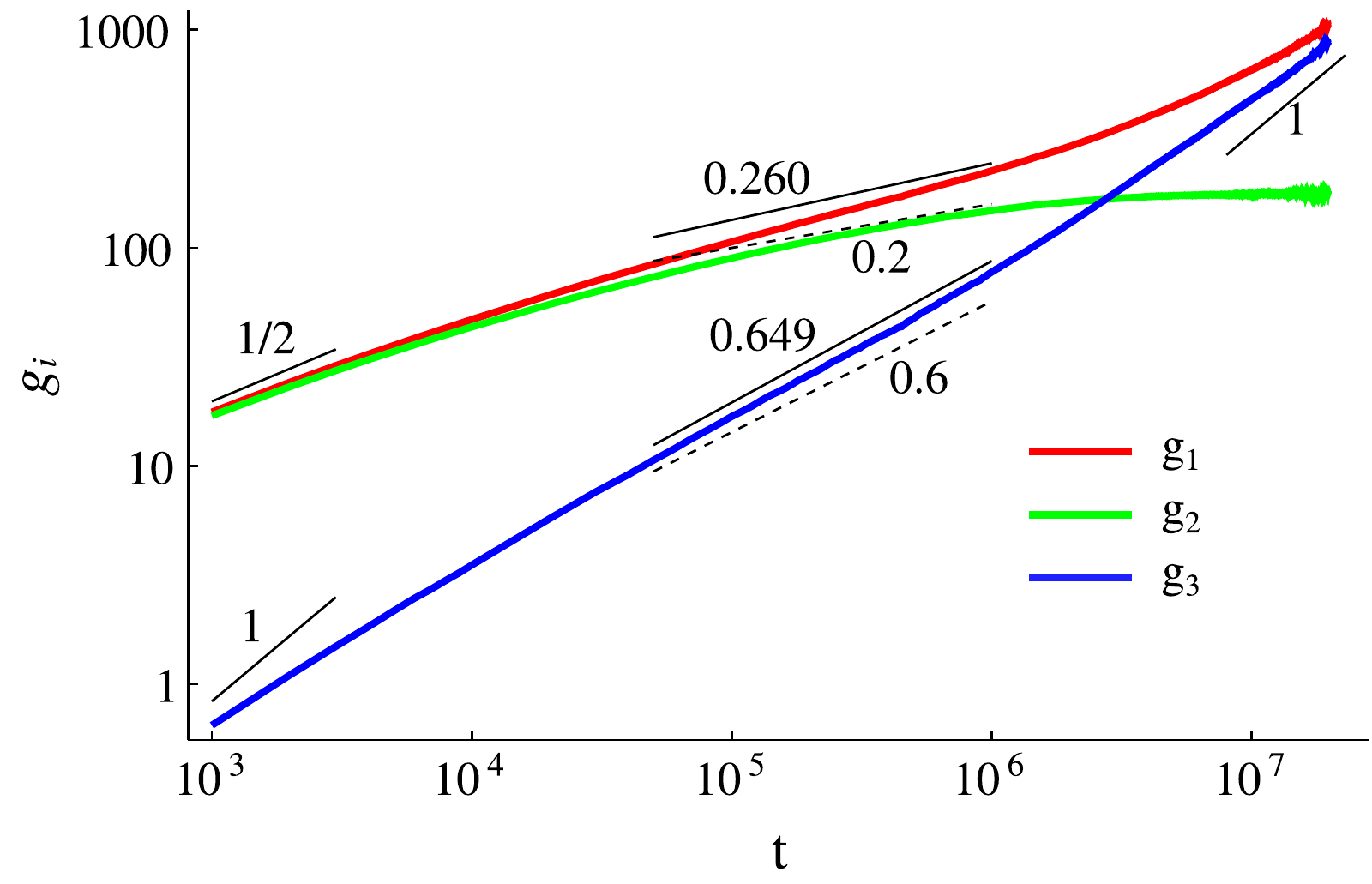}
 \caption{All three MSDs for rings of $N=800$ from simulations in \cite{Kremer_rings_dynamics} (thick red, green and blue lines). Slopes of thin black lines
(values indicated) represent exponents of our calculations - solid: annealed trees, dashed: ideal trees model.}
 \label{fig_gs}
\end{figure}

The time dependence of the mean square displacements exhibit the Rouse behavior at early times characterized by exponent $1/2$ for $g_{1}$ and $1$ for $g_{3}$
(Fig.~\ref{fig_gs}) . At intermediate times we found for $g_{1}$ exponent $2\nu/(2+\rho)$ which is about $0.260$ for the annealed trees and $0.2$ for the ideal
case. For $g_{3}$ the exponent is $(2\nu+1)/(\rho+2)$ that is $0.649$ for the annealed trees and $0.6$ for ideal ones. 

\begin{figure}[htb]
 \includegraphics[scale=0.5]{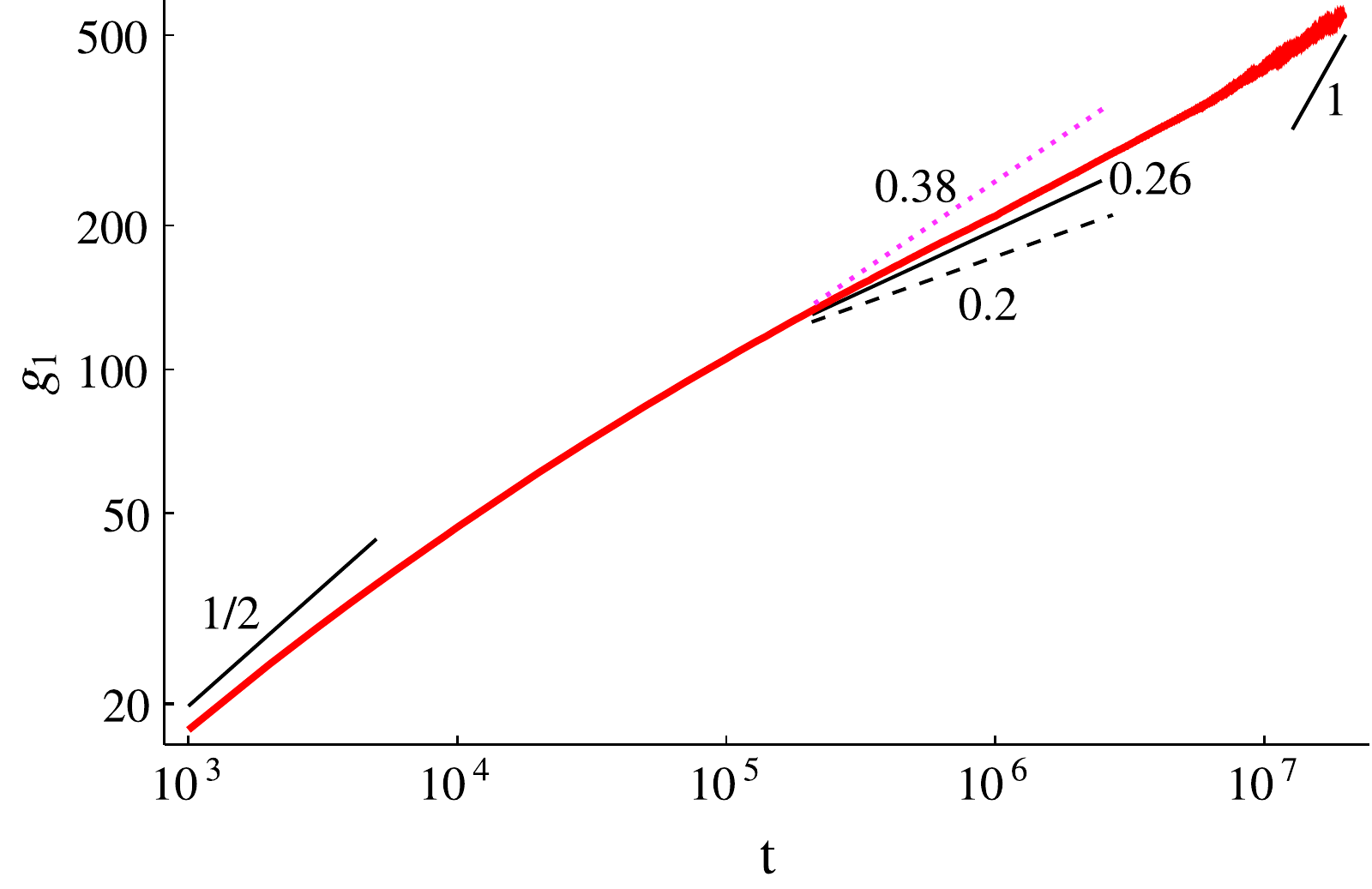}
 \caption{MSD $g_{1}$ for rings of $N=1600$ from simulations in \cite{Kremer_rings_dynamics} (thick red line).  Slope of the thin dotted pink line $0.38$
represents the exponent of the numerical result of \cite{Chertovich}. Slopes of thin black lines (values indicated) represent exponents of our calculations -
solid: annealed trees, dashed: ideal trees model. These rings have not fully reached diffusive regime (slope $1$ for long times) in the simulation time.}
 \label{fig_g1}
\end{figure}

\begin{figure}[!ht]
 \includegraphics[scale=0.5]{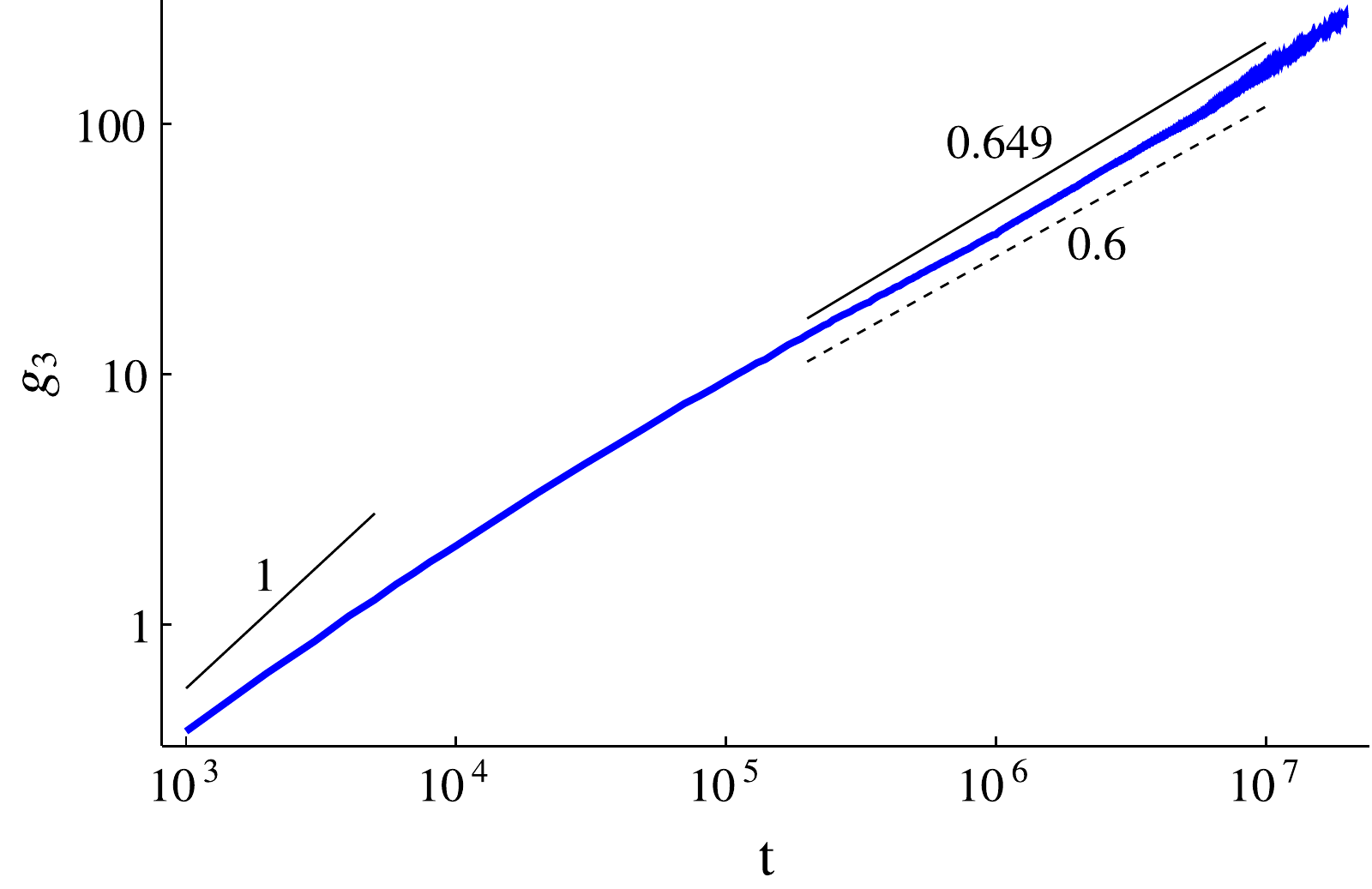}
 \caption{MSD $g_{3}$ for rings of $N=1600$ from simulations in \cite{Kremer_rings_dynamics} (thick blue line).  Slopes of thin black lines (values indicated)
represent exponents of our calculations - solid: annealed trees, dashed: ideal trees model. These rings have not fully reached diffusive regime (slope $1$ for
long times) in the simulation time.}
 \label{fig_g3}
\end{figure}

While $g_{3}$ agrees well at intermediate times (Fig.~\ref{fig_g3}), the exponent of $g_{1}$ (Fig.~\ref{fig_g1}) is found somewhat larger in the simulation
\cite{Kremer_rings_dynamics} (best fit gives about $0.3$, not shown). A reason for this numerical observation might be that for the simulated ring lengths the
intermediate regime is not long enough to capture the complete crossover to the calculated exponent. Nevertheless, numerically we clearly see a decrease of the
exponent at the intermediate timescales. Additional reason can be again the assumption that treats the topological obstacles as immobile. As noted above the
validity of this assumption at intermediate to late times is questionable as a significant rearrangement already started at the scale of branches i.e. the
scales larger than the lattice constant. This is also suggested by the numerical results of the internal rearrangement relaxation time. 

Another recent theoretical and numerical study \cite{Chertovich} investigated $g_{1}$ for long linear polymer in dense fractal conformation. Even though the
polymer has open ends it is prepared in a conformation with similar properties (gyration radius and contact probability scaling) to those measured for the rings
in melt. Then the mean squared displacement of the monomers is measured in the process of equilibration, where an exponent around $0.38$ at intermediate times
is found (indicated in Fig.~\ref{fig_g1}). The reason to compare this study with the present work and simulation \cite{Kremer_rings_dynamics} is the common
principle of topological constraints governing the genome folding and the melt of rings conformations. The different results from the two simulations deserve a
deeper study as they can help to understand the genome-ring melt correspondence. The analytical part of \cite{Chertovich} (predicting exponent about $0.4$)  and
the present work rely on different assumptions, but both take a mean-field approach. While \cite{Chertovich} models the effect of surrounding chains by
exploiting the chains surface, here we rely on the model of a ring as an annealed tree. Both mean-field approaches neglect density fluctuations that might have
an effect on the critical exponents as was demonstrated for the tube model in linear melts \cite{Semenov} and in the recently numerically studied ring in a gel
through the effect of self-threading \cite{Turner_ring_selfthreading_gel}. It is an open question to investigate the sensitivity of the critical exponents of
both the statics and dynamics of melt of rings beyond the mean-field considerations.
 
Both theories, of the ideal and the self-avoiding backbone agree very well with the experimental and numerical evidence. Unfortunately, the rather weak
dependence of the measurable exponents on $\rho$ in the investigated range does not allow to conclude in favor of any of the competing theories.

\section{Discussion}
\label{sec_discussion}
As we mentioned in the introduction above, melt of rings in general and its dynamics in particular is a challenging fundamental problem of polymer physics. At
the same time, as we also mentioned, this problem appears to have deep connections to genome 3D structure. Accordingly, here we discuss possible relevance of
our model findings for the genome folding field.

Experimentally, quite a few works investigated the 3D dynamics of chromatin in the cells nuclei of mammals
\cite{Garini_PhysRevLett_2009,Garini_PRE_2013,Garini_talk,Wirtz_2004,Wirtz_BioPhysJ_2011,Shiva_2012,Zidovska_PNAS_2013}, drosophila
\cite{Sedat_1997,Vazquez_drosophila_2001} and yeast \cite{Zimmer_yeast_dynamics_2006}; genome dynamics in bacterial nucleoid was studied as well
\cite{Theriot_Spakowitz_2010,Theriot_2012}. In most cases fluorescent microscopy was used. The common theme of all experimental results is a subdiffusive
behavior, typically with a power close to $0.4$: $g_{1}\sim t^{0.4}$. Such behavior is reported for both, parts of genome itself such as telomeres
\cite{Garini_PhysRevLett_2009,Garini_PRE_2013,Garini_talk} or certain gene loci \cite{Garini_talk,Zimmer_yeast_dynamics_2006}, as well as foreign nanoparticles
immersed into the nucleus \cite{Vazquez_drosophila_2001, Wirtz_2004, Wirtz_BioPhysJ_2011, Shiva_2012}. Typically subdiffusive motion is reported over a limited
interval of time --
 such that overall displacement during this time is pretty small, usually only a few tens of nanometers, well below $0.1 \unit{\mu m}$.

Although observation of sub-diffusion is encouraging in terms of application of polymer models with topological constraints, the detailed quantitative
comparison is certainly premature at this stage. The main reason is the fact that genome dynamics in the cell is active, driven, and ATP dependent
\cite{RevModPhys_ActiveSoftMatter_2013,Zidovska_PNAS_2013}. Active hydrodynamics of chromatin was recently considered in \cite{Grosberg_Zidovska_2014}. One
interesting observation (see specifically Fig.~3 in the work \cite{Grosberg_Zidovska_2014}) is that chromatin dynamics in active and starved cells appears to be
different only on the length scales in excess of one micron or so. This may give a credence for more detailed association of subdiffusive behavior in folded
genomes and in topologically restricted polymer models. On a more cautious note, the goal of our study of polymer dynamics should be to understand their
rheological behavior in terms of moduli dependence on both frequency and wave vector -- the information that must be fed into the phenomenological theory of
chromatin hydrodynamics \cite{Grosberg_Zidovska_2014}. Our present work should be viewed as a step in that direction.

\section*{Acknowledgement}
This work was supported in part by the National Science Foundation under Grant No. PHYS-1066293 and AYG acknowledges the hospitality of the Aspen Center for
Physics. The authors thank Michael Rubinstein for useful and stimulating discussions before, during, and after his talk \cite{Rubinstein_talk}.

\appendix
\section{Gyration radius distribution for a tree and backbone definition -- a generalization of Kramers theorem}
\label{app_L_definition}

Flory-type theory for rings \cite{Grosberg_ring_melt_static} assumes that ring conformation is controlled by the balance of two competing entropic effects. On
the one hand approaching the doubly folded conformation makes it easy to penetrate other rings, which is entropically favorable. On the other hand, double
folding the ring is entropically unfavorable. The former factor is described by free energy $\sim R^{2}/L$, the latter $\sim L^{2}/N$; their balance is in the
heart of the Flory theory \cite{Grosberg_ring_melt_static}. To understand it better, including the meaning of $L$, we here derive the $R^{2}/L$ result in
greater detail. 

As usually in Flory theory, the $R^{2}/L$ free energy should be viewed as entropy price of deforming a Gaussian polymer, in this case -- a tree with quenched
branches.

The entropic price for swelling a tree can be calculated from probability distribution of gyration radius $R$, particularly the tail of the distribution when
$R$ is large.
As will be shown below, this also defines the backbone $L$. The approach is modeled after the work \cite{Fixman_gyration_radius} (with small corrections in
\cite{Forsman_gyration_radius_correction,Fujita_gyration_radius_correction}) and \cite{Moore_Lua_Grosberg_knotted_polymers}
(\cite{Fixman_gyration_radius,Forsman_gyration_radius_correction,Fujita_gyration_radius_correction} considered probability distribution of gyration radius for
linear Gaussian polymer; \cite{Moore_Lua_Grosberg_knotted_polymers} did the same for a phantom Gaussian ring). 

\begin{figure}[htb]
 \includegraphics[scale=1]{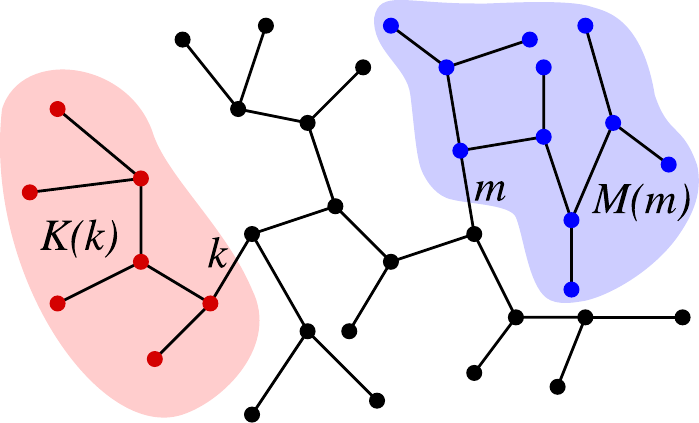}
 \caption{Towards the definition of matrix $G(k,m)$. On the tree, bonds $k$ and $m$ naturally divide all nodes of the graph into three categories: on one side
of $k$ (shaded), on the other side of $m$ (also shaded), and in between (not shaded).  We denote the numbers of monomers in the former two categories as $K(k)$
and $M(m)$. Each matrix element is given by $G(k,m)=K(k)M(m)/N^{2}$. In this particular example $G(k,m)=7\times 11/36^{2}$.}
 \label{fig_Gmatrix}
\end{figure}

Suppose for simplicity that functionality of branch points in the tree is just $z=3$, and there are only branch points and ends, namely, $n$ branch points and
$n+2$ ends, i.e. a total of $N=2n+2$ monomers. If $\vec{r}_{i}$ are position vectors of these ``monomers'' then gyration radius reads
\begin{equation}
 R^{2} = \frac{1}{2N^{2}} \sum_{i,j} (\vec{r}_{i}-\vec{r}_{j})^{2}.
\end{equation}
On the tree, every $\vec{r}_{i}-\vec{r}_{j}$ is uniquely represented as the sum of the proper set of bond vectors
$\vec{\eta}_{k}=\vec{r}_{i^{\prime}}-\vec{r}_{i}$, where $i$ and $i^{\prime}$ are monomers connected by bond $k$.
Accordingly, gyration radius can be represented as 
\begin{equation}
\label{RgG}
 R^{2} = \sum_{k,m} G(k,m) \vec{\eta}_{k}\vec{\eta}_{m},
\end{equation}
where the indices $k$ and $m$ label bonds (unlike $i$, $i^{\prime}$ and $j$ above, which label nodes or monomers), and $G(k,m)$ is an interesting $N-1$ by $N-1$
matrix, illustrated in Fig.~\ref{fig_Gmatrix}: each matrix element is given by $G(k,m) = K(k) M(m)/N^{2}$, where $K(k)$ is the number of monomers on one side of
bond $k$, while $M(m)$ is similarly the number of monomers on the other side of bond $m$. 

For a Gaussian system (without excluded volume and when ``bonds'' are long enough compared to persistence length), each bond vector $\vec{\eta}$ is Gauss
distributed, and vectors $\vec{\eta}_k$ and $\vec{\eta}_m$, with $k \neq m$, are independent. Accordingly, formula (\ref{RgG}) returns the known result for
mean squared average gyration radius of the tree, it involves only diagonal elements $G(k,k)$ and is given by the trace of that matrix:
\begin{equation}
\label{Gkk}
 \langle R^{2} \rangle = \sum_{k} G(k,k) = a^{2} \sum_{k=1}^{N-1} \frac{K(k)(N-K(k))}{N^{2}},
\end{equation}
which is Kramers theorem \cite{Kramers} that can be found also in e.g. \cite{Rubinstein_book}; here $a^2 = \left< \vec{\eta}^2 \right>$ is the mean squared bond
length.

For a Gaussian tree we can do more and find not only averaged $R^2$, but the probability distribution of $R^2$, since each bond has the probability
distribution $\sim e^{-3\vec{\eta}^{2}/2a^{2}}$. The characteristic function of $R^{2}$ reads
\begin{equation}
\Phi(s) = \langle e^{i s R^{2}}\rangle = A \int e^{-\sum_{k}\frac{3\vec{\eta}_{k}^{2}}{2a^{2}} + i s \sum_{k,m}
G(k,m)\vec{\eta}_{k}\vec{\eta}_{m}} d^{3}\left\{\vec{\eta}\right\},
\end{equation}
where the explicit expression for normalization factor $A$ is dropped for brevity. Rotating now the coordinate system in this $N-1$  dimensional space of
$\left\{\vec{\eta}\right\}$ to the basis of eigenvectors $\{\vec{\xi}\}$ of matrix $G$, we obtain
\begin{equation}
\Phi(s) = A\int e^{-\sum_{p}\frac{3\vec{\xi}_{p}^{2}}{2a^{2}}(1-i s \lambda_{p})} d^{3}\{\vec{\xi}\} = \prod_{p=1}^{N-1} (1-i s \lambda_{p})^{-3/2},
\end{equation}
where $\lambda_{p}$ are eigenvalues. From here, finding the probability distribution of $R^{2}$ is a matter of inverse Fourier transform of $\Phi(s)$
\begin{equation}
 P(R^{2}) = \frac{1}{2\pi} \int e^{-i s R^{2}- \frac{3}{2}\sum_{p}\ln(1-i s \lambda_{p})}  ds.
\end{equation}
Since we are interested in the behavior of $P(R^{2})$ at large $R$ the asymptotics is controlled by the singularity of $\Phi(s)$ which is closest to the origin
in complex $s$-plane, that corresponds to the largest eigenvalue $max(\lambda_{p})=\lambda$. In the vicinity of this singularity there is a saddle point which
dominates the integral, and we can evaluate inverse Fourier transform integral by steepest descent. The equation for saddle point location reads $R^{2}\simeq
\frac{3}{2}\frac{\lambda}{1-i s\lambda}$. From here the result of saddle point integration (up to logarithmic corrections) is
\begin{equation}
\label{P}
 P(R^{2})|_{R\to\infty} \sim e^{-\frac{R^{2}}{\lambda} + \frac{3}{2}\ln\frac{2R^{2}}{3\lambda}} \sim e^{-R^{2}/\lambda} = e^{-c R^{2}/\langle R^{2}\rangle}.
\end{equation}
It is difficult to find the maximal eigenvalue $\lambda$ in general for arbitrary tree configuration, but looking at the equation (\ref{P}) it is natural to
imply that $\lambda$ is the mean squared average gyration radius up to a numerical factor $c$; this is what we did in the last step in formula (\ref{P}). Here
as above, mean squared average $\langle R^{2} \rangle$ is the thermal average, given by formula (\ref{Gkk}), for the given tree i.e. it does not involve
average over different tree topologies. Assuming that numerical coefficient $c$, which is a property of any given tree, is of order unity, we drop it for the purposes of scaling analysis in the
framework of Flory theory.

The entropic price for swelling the tree is hence given by 
\begin{equation}
\label{entropy_of_swelling}
\Delta S(R) =-c R^{2}/\langle R^{2}\rangle = - R^{2}/ a L .
\end{equation}
Since each end-to-end arm in the tree is Gaussian linear polymer, the latter relation says that $L$ is a properly (and non-trivially) averaged length of all
such arms. In other words we can define $L$ in terms of eigenvalue $\lambda$: $L = \lambda/a$. Although the result $R^{2}/L$ for the swelling entropy was known long ago
\cite{Daoud_Pincus_1983}, we should emphasize its non-trivial character because when the tree swells then all of its branches swell, not just one.

The result (\ref{P}) or (\ref{entropy_of_swelling}) is the generalization of Kramers theorem \cite{Kramers,Rubinstein_book}.

\section{Unconcatenated rings in 2D}
\label{app_mapping}
In this Appendix, we discuss the statistical equilibrium properties of the unconcatenated ring melt in 2D. The goal is to understand better the underlying
annealed tree model, its accuracy and limitations.

In 2D, instead of unconcatenated rings we should think of a system of rings such that no ring can be inside another ring. For such system, the double-folded
conformation is not a hypothesis nor approximation, but an obvious fact. Furthermore the resulting trees are definitely annealed. Flory theory
\cite{Grosberg_ring_melt_static} is generalized straightforwardly for this case yielding $R\sim N^{1/2}$ and $L\sim N^{\rho}$ with
$\rho=1/2\nu_{\mathrm{F_{2D}}}$, where $\nu_{\mathrm{F_{2D}}} = 3/4$ is the Flory exponent for regular self-avoiding walk in 2D. In reality as we explain below
based on the works \cite{Strasbourggroup_linear2Dmelt_2009, Duplantier_JSP_1989, Duplantier_LERW_1992, Lawler_SLE_UST_2004, Duplantier_review_2003}
$\rho=1/2\nu_{\mathrm{LERW}}$, where $\nu_{\mathrm{LERW}}=4/5$ is the exponent of the loop-erased random walk. If true, this indicates the Flory theory in this
case is not exact, but pretty close.
 
The fact that $\rho=1/2\nu_{\mathrm{LERW}}$ in 2D  means that the backbone of the ring is a loop-erased random walk with gyration radius $R\sim L^{\nu_{\mathrm{LERW}}}$.
We will demonstrate this by mapping a melt of linear chains onto a melt of rings and extract the statistics of the backbone of the underlying trees. Numerical
simulation \cite{Strasbourggroup_linear2Dmelt_2009} and exact calculations \cite{Duplantier_JSP_1989} of 2D melt of \emph{linear} chains show, that chains adopt
compact conformation with \emph{end-to-end distance} (that is equal to the gyration radius up to a numerical factor) $R\sim N^{1/2} \sim P^{4/5}$ where $P$ is
the perimeter of the chain defined as the number of monomers neighboring with a monomer of other chain. For simplicity, consider the linear melt system on a
lattice and construct the corresponding dual lattice. Let us consider two neighboring chains and construct a path on the dual lattice separating the two chains.
This path of length proportional to $P$ has the statistics of loop-erased random walk $R\sim P^{4/5}$ and forms a backbone of a tree. The branches of the tree
spread from this backbone to the regions occupied by the two chains 
on the dual lattice such that they are not allowed to cross the chains. This way we construct a tree with $R\sim L^{\nu_{\mathrm{LERW}}}$. This procedure can be repeated for every pair of neighboring chains.

This way we map the melt of linear chains in 2D onto the melt of annealed trees in 2D. By circumscribing the ring around each tree, we can construct also the
melt of unconcatenated rings. Thus we obtained one-to-one mapping between all three systems - melt of unconcatenated rings, melt of linear chains and melt of
annealed trees. All of them are, therefore, described by the same index $\nu_{\mathrm{LERW}}$.

Additionally, the same index describes also uniform spanning tree \cite{Lawler_SLE_UST_2004} (and, therefore, single linear chain or single loop) densely
confined inside a region with smooth boundaries .

Similar problem of a single unknotted loop in a cavity in 3D was recently studied by simulations \cite{Mirny_one_ring_2014}, the analogy with the melt of rings
is confirmed, but perhaps not to the extent of exact mapping like in 2D.

\bibliography{RingDynamics_References}
\bibliographystyle{naturemag}
\end{document}